\documentclass[a4paper]{article}
\usepackage[cp1250]{inputenc}
\usepackage[T1]{fontenc}
\usepackage[pdftex]{graphicx}
\newcommand{\la}{\lambda_1}

\newcommand{\lc}{\lambda_3}
\newcommand{\ld}{\lambda_4}
\newcommand{\lp}{\lambda_5}

\newcommand{\lczp}{\lambda_{345}}
\newcommand{\ba}{\begin{array}}
\newcommand{\ea}{\end{array}}
\newcommand{\g}{\,\mathrm{GeV}}
\newcommand{\m}{M_{H^{\pm}}}
\newcommand{\rg}{R_{\gamma\gamma}}
\newcommand{\rzg}{R_{Z\gamma}}
\usepackage{amsmath}
\usepackage{amssymb}
\usepackage{amsthm}
\usepackage{amsfonts}
\usepackage{cite}
\begin{document}
\author{M.~Krawczyk, D.~Soko\l owska, P.~Swaczyna, B.~\'{S}wie\.{z}ewska\\
\small{\textit{Faculty of Physics, University of Warsaw, Ho\.{z}a 69, 00-681 Warsaw, Poland}}
}
\title{Higgs $\to \gamma\gamma$, $Z\gamma$ in the Inert Doublet Model
\thanks{Presented at Matter To The Deepest,
Recent Developments In Physics Of Fundamental Interactions,
XXXVII International Conference of Theoretical Physics by~Bogumi\l a \'{S}wie\.{z}ewska}%
}
\maketitle
\begin{abstract}
An analysis of the Higgs boson decay rates to $\gamma\gamma$ and $Z\gamma$ in the Inert Doublet Model is presented. We study the correlation between the two rates and perform extended analysis of the two-photon rate ($\rg$). We study both the possibility of enhancing and suppressing $\rg$ and find constraints for masses of the scalar particles (in particular the dark matter (DM) candidate and the charged scalar) and their couplings to the Higgs boson.  We also combine the resulting constraints with those following from the WMAP measurements of the DM relic density, obtaining stringent constraints on different dark matter scenarios.
\end{abstract}

  
\section{Introduction}
The decay channel of the Higgs boson to two photons is one of the most important channels for the Higgs searches at the LHC. The current measurements report on the following values of the signal strength: $\rg = 1.65\pm0.24\mathrm{(stat)}^{+0.25}_{-0.18}\mathrm{(syst)}$ (ATLAS)~\cite{ATLAS:2013oma}, $\rg =0.79^{+0.28}_{-0.26}$ (CMS)~\cite{CMStalk}. They are both consistent with the SM prediction ($\rg=1$), but do not exclude the possibility of deviations due to new physics (NP) contributions. For the $h\to Z\gamma$ channel there are still not enough data to draw significant conclusions.

The loop-induced decays of the Higgs boson are important for the NP searches, because in this type of processes the contributions from the new particles can be comparable to the SM ones. In the Inert Doublet Model (IDM) the $h\to gg$ width is not modified with respect to the SM, while the $h\to\gamma\gamma$ and $h\to Z\gamma$ can receive corrections due to the $H^{\pm}$ exchanged in a loop. Also, the invisible Higgs decays can modify the signal strengths, augmenting the total width of the Higgs boson. So the $h\to\gamma\gamma$ channel is a perfect one to reveal some information about the extra scalars present in the IDM~\cite{rg, Goudelis:2013, jhep}. As the experimental results are not conclusive yet, we analyze both the case of enhancing and  suppressing, as compared to the SM, the two-photon decay rate.


\section{Inert Doublet Model} \setlength\arraycolsep{2pt}
The IDM is a Two Higgs Doublet Model with the scalar fields $\phi_S$ and $\phi_D$ interacting according to the following scalar potential~\cite{Ma:1978, Cao:2007rm,Barbieri:2006dq, Krawczyk:2010}
\begin{eqnarray}
V&=&-\frac{1}{2}\left[m_{11}^{2}(\phi_{S}^{\dagger}\phi_{S})+m_{22}^{2}(\phi_{D}^{\dagger}\phi_{D})\right]+\frac{1}{2}\left[\lambda_{1}(\phi_{S}^{\dagger}\phi_{S})^{2}+\lambda_{2}(\phi_{D}^{\dagger}\phi_{D})^{2}\right]\nonumber\\*
&&+\lambda_{3}(\phi_{S}^{\dagger}\phi_{S})(\phi_{D}^{\dagger}\phi_{D})+\lambda_{4}(\phi_{S}^{\dagger}\phi_{D})(\phi_{D}^{\dagger}\phi_{S})+\frac{1}{2}\lambda_{5}\left[(\phi_{S}^{\dagger}\phi_{D})^{2}+(\phi_{D}^{\dagger}\phi_{S})^{2}\right],\nonumber
\end{eqnarray}
where all the parameters are real and $\lp<0$. We set the Yukawa interactions to Type I, i.e., only the $\phi_S$ doublet couples to fermions. The vacuum state is such that only $\phi_S$ develops a non-zero vacuum expectation value (VEV): \renewcommand{\arraystretch}{.7}$\langle\phi_{S}\rangle=\left(\begin{array}{c}0\\v/\sqrt{2}\\\end{array}
\right)$, $\langle\phi_{D}\rangle=\left(\begin{array}{c}0\\0\\\end{array}
\right)$. 

The mass matrix written in terms of the component fields of $\phi_S$ and $\phi_D$ is diagonal, so the two doublets do not mix. The particle spectrum of the IDM consists of a SM-like Higgs boson $h$ which originates from $\phi_S$ and thus couples to fermions and gauge bosons just like the SM Higgs boson, and four so-called dark (or inert) scalars which originate from $\phi_D$ and do not couple to fermions at the tree level. The masses of the particles read
\begin{eqnarray}
&&M_{h}^{2}=m_{11}^{2}=\la v^{2},\quad M_{H^{\pm}}^{2}=\frac{1}{2}(\lc v^{2}-m_{22}^{2}),\nonumber\\* &&M_{A}^{2}=\frac{1}{2}(\lczp^{-}v^{2}-m_{22}^{2}),
\quad M_{H}^{2}=\frac{1}{2}(\lczp v^{2}-m_{22}^{2}),\nonumber
\end{eqnarray}
where $\lczp=\lc+\ld+\lp,\;\lczp^-=\lc+\ld-\lp$.

The model (the Lagrangian and the vacuum state) is exactly symmetric under a $D$ symmetry, such that: $\phi_D \xrightarrow{D} -\phi_D$, $\phi_S \xrightarrow{D} \phi_S$, $\phi_{\mathrm{SM}} \xrightarrow{D} \phi_{\mathrm{SM}}$. Due to the $D$-parity conservation, the lightest $D$-odd particle is stable and has been shown in Refs.~\cite{Dolle:2009fn} to provide a viable dark matter (DM) candidate. In this work we choose $H$ to be the lightest among the dark scalars. The DM particle of mass in three regions agrees with the WMAP measurements of the DM relic density ($0.1018<\Omega_{DM} h^2<0.1234$ at $3\sigma$~\cite{wmap-results}): light DM ($M_H\lesssim 10\g$), intermediate DM ($40\g\lesssim M_H\lesssim 150\g$) and heavy DM ($M_H>500\g$).

The model is in agreement with current experimental and theoretical constraints: positivity of the potential, perturbative unitarity, stability of the Inert vacuum, electroweak precision tests and the LEP constraints on the scalars' masses, see e.g. Ref.~\cite{Swiezewska:2012}. As $H$ is the DM candidate it has to be the lightest of the dark scalars, i.e. $M_H<M_A$, $\m$. Moreover, we set the mass of the SM-like Higgs boson to $M_h=125\g$.

\section{$\gamma\gamma$ and $Z\gamma$ decay rates of the Higgs boson}
The $\gamma\gamma$  decay rate of the Higgs boson is defined as follows\footnote{The $Z\gamma$ decay rate is defined analogously.}
$$
R_{\gamma \gamma}:=\frac{\sigma(pp\to h\to \gamma\gamma)^{\textrm{IDM}}}{\sigma(pp\to h\to \gamma\gamma)^{\textrm  {SM}}}\approx\frac{\textrm{Br}(h\to\gamma\gamma)^{\textrm {IDM}}}{\textrm{Br}(h\to\gamma\gamma)^{\textrm {SM}}},
$$
where the narrow width approximation has been used. Moreover, the facts that the main production channel of the Higgs boson is the gluon fusion, and that the cross section for this process in the IDM is the same as in the SM have been taken into account.

\begin{figure}[ht]
\centering
\includegraphics[width=5.2cm]{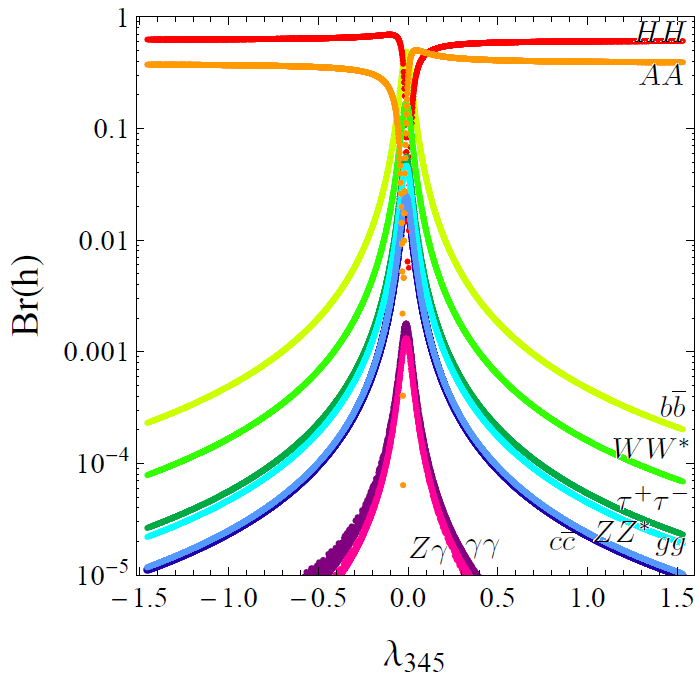}\hspace{.5cm}
\includegraphics[width=5.2cm]{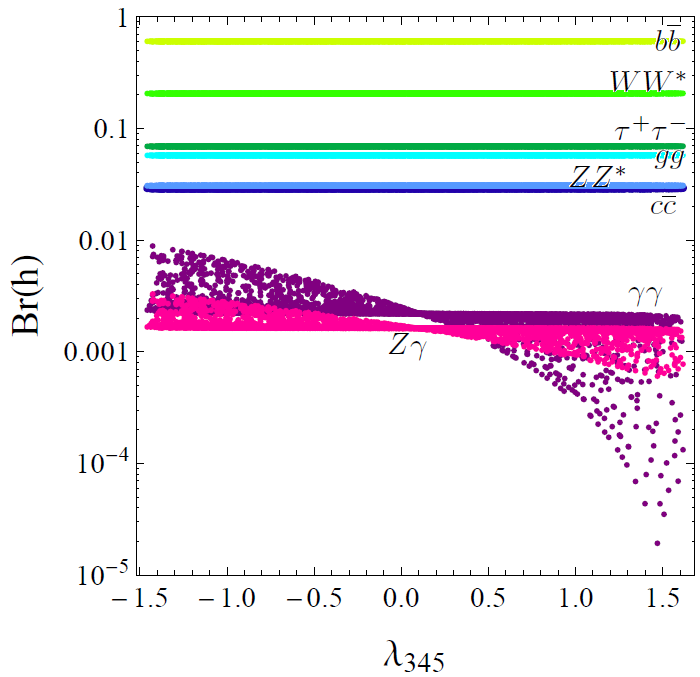}
\caption{Branching ratios of the Higgs boson as functions of $\lczp$. Left panel: Invisible channels are open ($M_H=50\g$, $M_A=58\g$). Right panel: Invisible channels are closed ($M_H=75\g$, $M_A>M_H$).
\label{br}}
\end{figure}

This rate can be modified with respect to the SM prediction ($\rg=1$) in two ways. First, the invisible decay channels ($h\to HH$, $h\to AA$) can augment the total decay width of the Higgs boson $\Gamma(h)$. Their widths are controlled by the masses of the neutral dark scalars and their couplings to~$h$: $\lambda_{345}\sim hHH$ and $\lambda_{345}^-\sim hAA$.  Secondly, the charged scalar can be exchanged in loops leading to modification of the partial decay width of the Higgs boson to two photons, $\Gamma(h\to \gamma\gamma)$. The $H^{\pm}$ contribution is controlled by $m_{22}^2$ and $\m$ (alternatively $\lambda_3\sim hH^+H^-$ and $\m$ can be used).\footnote{The same applies for $\rzg$.} In Fig.~\ref{br} the branching ratios of the Higgs boson as functions of $\lczp$ are presented, for the case when the invisible channels are open (left panel) and closed (right panel). It can be seen that while the invisible channels are kinematically allowed, they dominate over the SM channels and once they are closed, the impact of the charged scalar loop on $h\to \gamma\gamma$ and $h\to Z\gamma$ becomes visible.
\subsection{Correlation between $\rg$ and $\rzg$}
\begin{figure}[ht]
\centering
\includegraphics[width=0.5\textwidth]{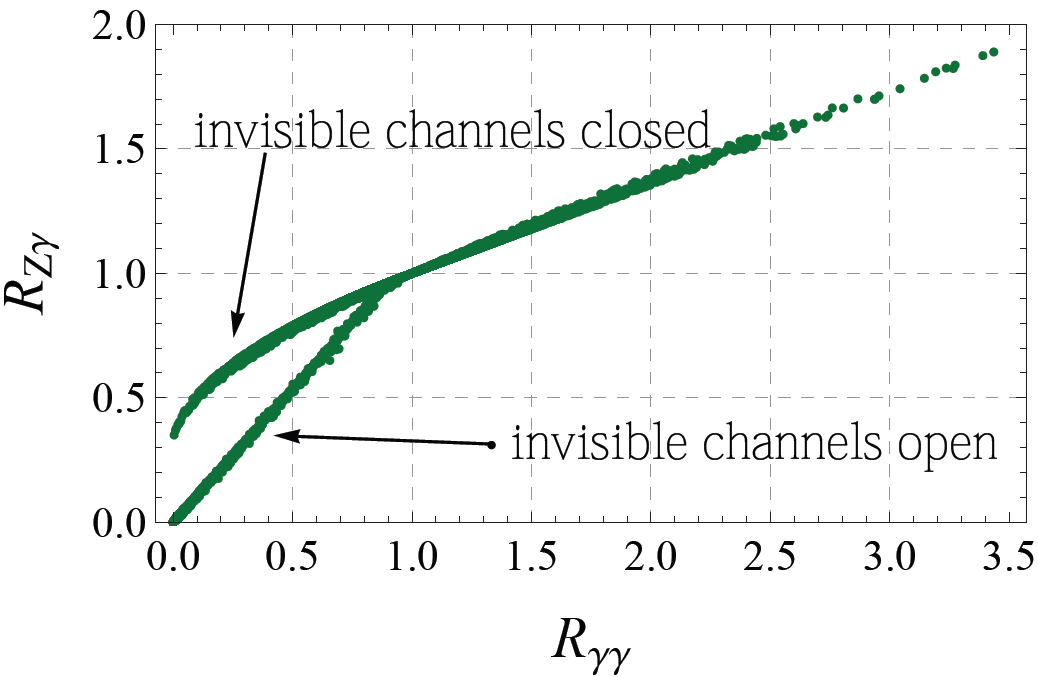}
\caption{Correlation between $\rg$ and $\rzg$.
\label{korelacja}}
\end{figure}

The correlation between $\rg$ and $\rzg$ is presented in Fig.~\ref{korelacja}~\cite{rg}. When the invisible channels are open, both $\rg$ and $\rzg$ are suppressed, as they are damped by the invisible decays width. The lower branch of the curve from Fig.~\ref{korelacja}, where $\rg\approx\rzg<1$ corresponds to this case. When the invisible decay channels are closed, the dependence of $\rg$ and $\rzg$ on the $H^{\pm}$ loop is visible, see the upper branch of the curve in Fig.~\ref{korelacja}.
\section{Enhanced two-photon rate}
\begin{figure}[ht]
\centering
\includegraphics[width=5.5cm]{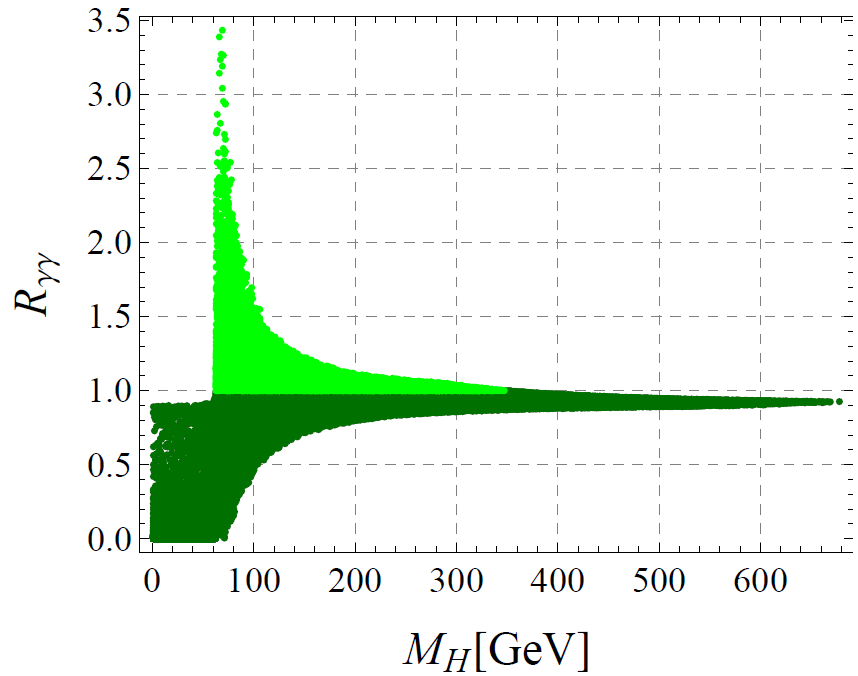}\hspace{.5cm}
\includegraphics[width=5.5cm]{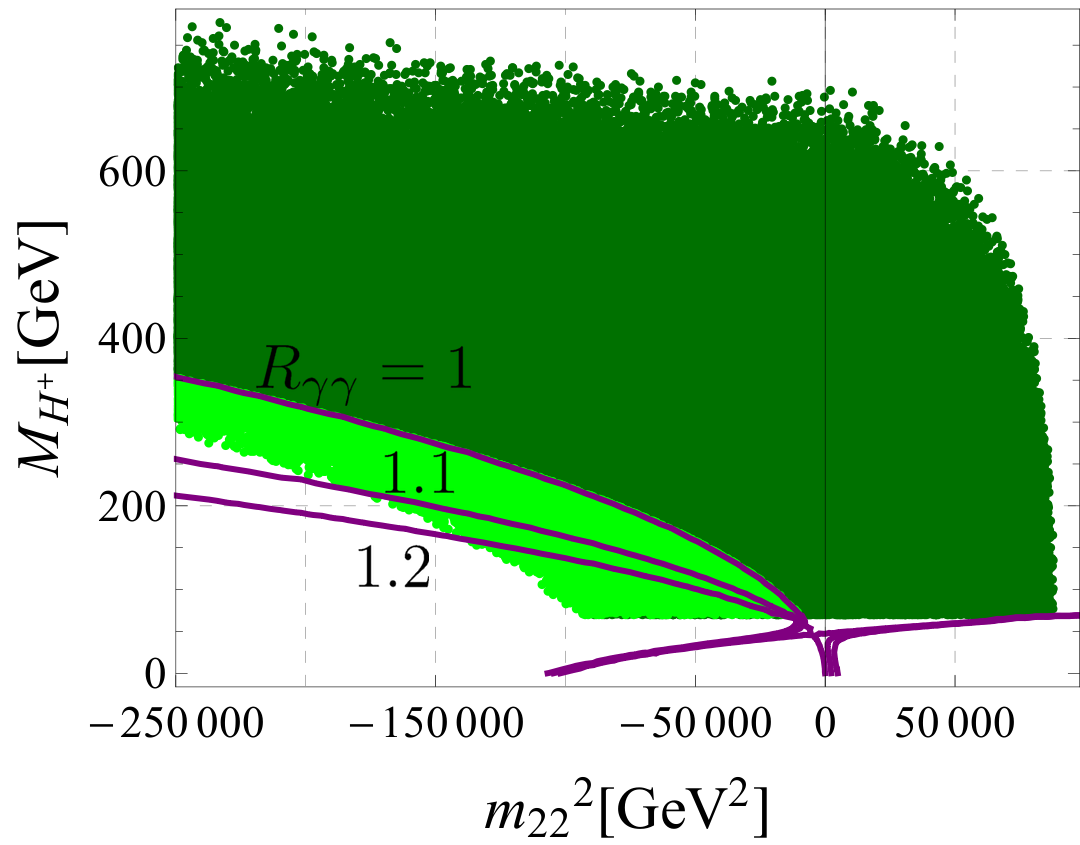}
\caption{Left panel: $\rg$ as a function of $M_H$. Right panel: Region allowed by the experimental and theoretical constraints in the $(m_{22}^2,\ \m)$ plane. Light green (gray) indicates the region where $\rg\geqslant1$, the lines correspond to the constant values of $\rg$. Plots are made for $-25\cdot10^4 \g^2\leqslant m_{22}^2\leqslant 9\cdot 10^4\g^2$.
\label{enhanced}}
\end{figure}

The dependence of $\rg$ on $M_H$ (Fig.~\ref{enhanced}, left panel) shows that $\rg>1$ is not possible if $M_H<M_h/2$. This means that if enhanced two-photon rate is observed, then the light DM candidate is excluded, and the heavier dark scalars also have to be above the kinematical threshold for the $h\to\mathrm{scalars}$ decay, i.e., $M_A$, $\m>M_h/2$.

In the right panel of Fig.~\ref{enhanced} the viable parameter region in the $(m_{22}^2,\ \m)$ plane is presented. The region where $\rg>1$ is not bounded, so enhancement of $\rg$ is possible even for very heavy $H^{\pm}$ (provided that $m_{22}^2$ is also big and negative). However, regions where substantial enhancement takes place are bounded, for example assuming $\rg>1.2$ implies $\m\lesssim154\g$ and thus $M_H\lesssim154\g$ as well. Combining this with the previously described results and LEP bounds on $\m$ gives stringent bounds on $\m$, $M_H$~\cite{rg}:
$$
62.5\g<M_H\lesssim154\g,\quad 70\g<\m\lesssim154\g.
$$

\section{Suppressed two-photon rate}
If $\rg<1$ is considered, constraints on the parameter space of the IDM can be found as well~\cite{jhep}. Below we present results following from the requirement that $\rg>0.7$, as suggested by the CMS data.

\begin{figure}[ht]
\centering
\includegraphics[width=5cm]{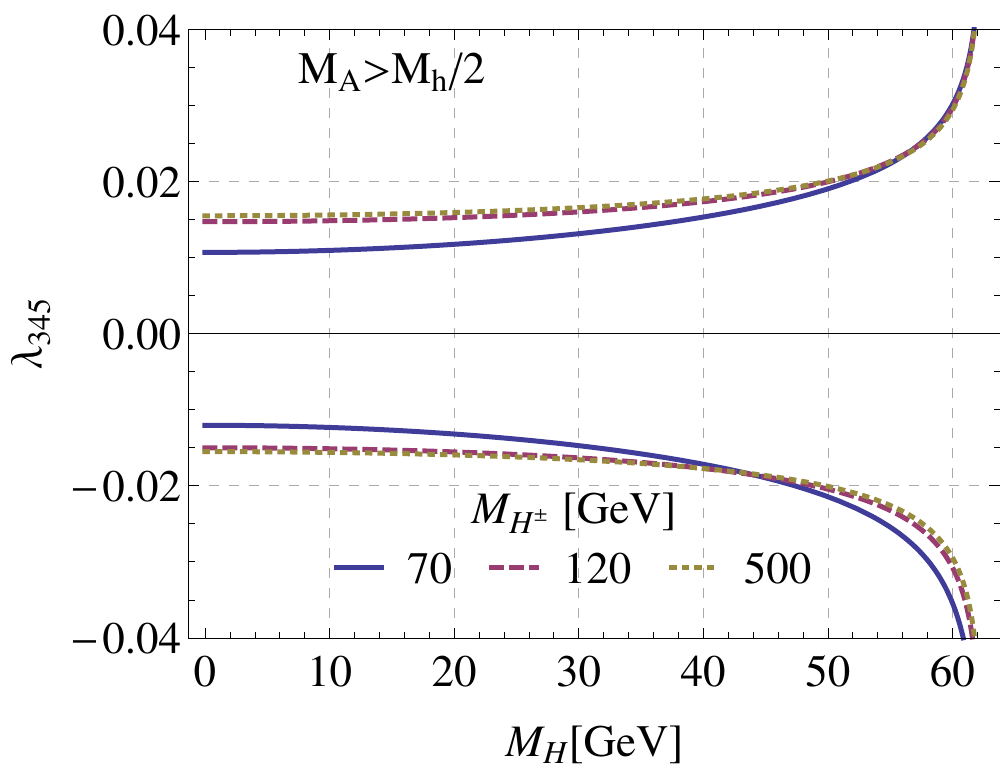}\hspace{.5cm}
\includegraphics[width=5cm]{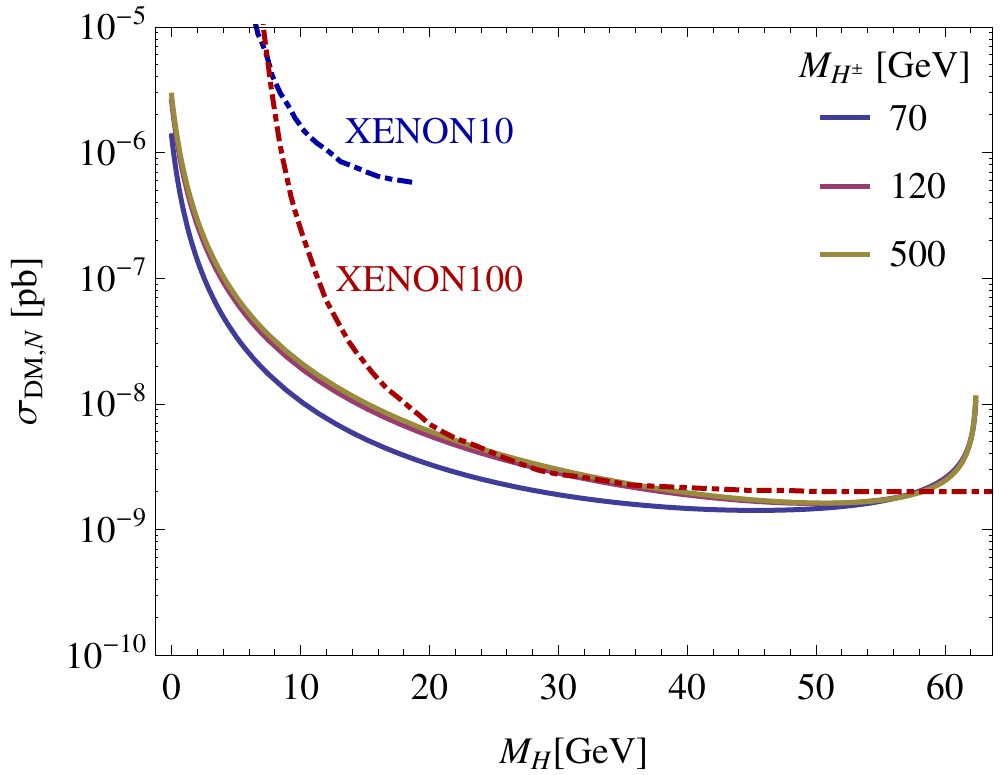}
\caption{Left panel: Constraints on $\lczp$ coming from $\rg>0.7$ as functions of~$M_H$. Right panel: Constraints in the $(M_H,\sigma_{DM,N})$ plane. The figures come from Ref.~\cite{jhep}.
\label{lczp constraints}}
\end{figure}

Once masses of the dark scalars are fixed, $\rg$ can be expressed as a function of $\lczp$. The function $\rg(\lczp)$ is bell-shaped, so setting a lower limit on $\rg$ implies upper and lower bounds on $\lczp$. These bounds of course depend on the masses of the dark scalars. The upper and lower bounds on $\lczp$ as functions of $M_H$, for different values of $\m$, are presented in the left panel of Fig.~\ref{lczp constraints}. They can be translated to the $(M_H,\sigma_{DM,N})$ plane ($\sigma_{DM,N}\sim \lczp^2$) and compared with the constraints set by the XENON100 results, see the right panel of Fig.~\ref{lczp constraints}. As can be seen, the $\rg$ constraints (solid lines) are stronger or comparable to the WMAP bounds (dashed lines). Also a comparison with the constraints following from the measurement of invisible decays branching ratios at the LHC (Br$(h\to\mathrm{inv})<65\%$) can be made, see~\cite{jhep}.
\subsection{Combination with the WMAP constraints}

\begin{figure}[ht]
\centering
\includegraphics[width=0.33\textwidth]{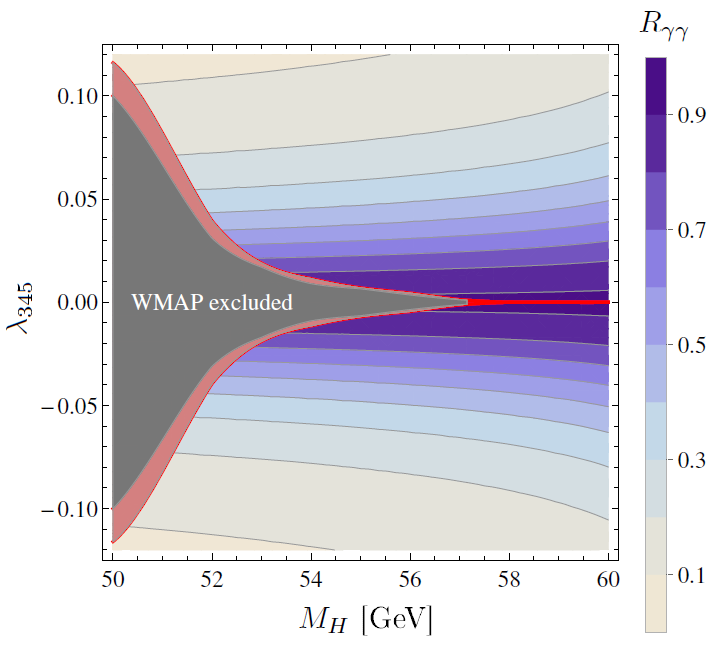}
\includegraphics[width=0.33\textwidth]{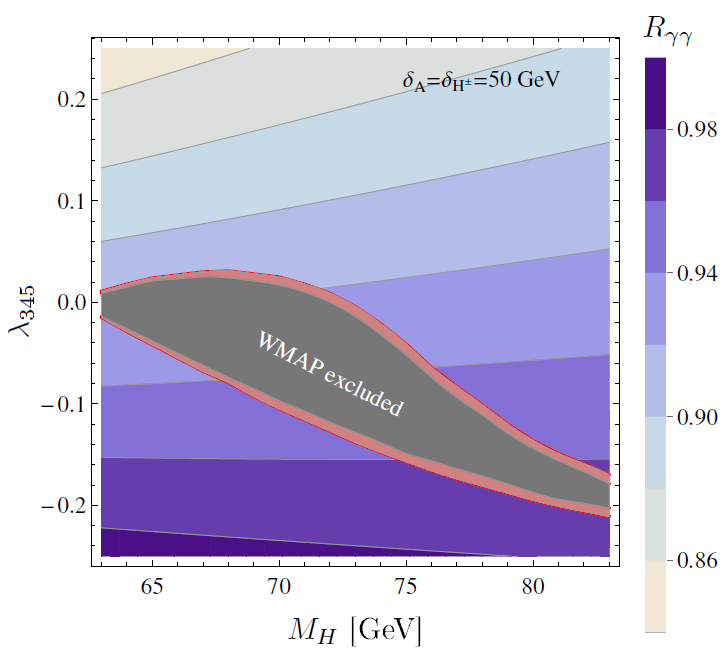}
\includegraphics[width=0.33\textwidth]{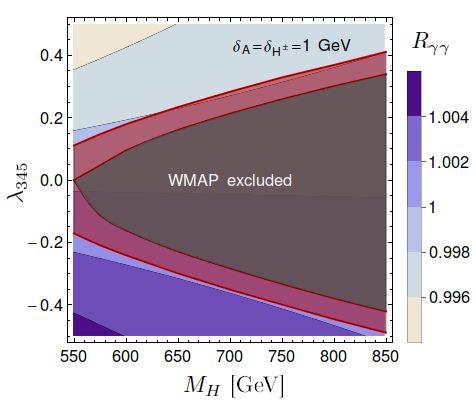}
\caption{Constraints following from WMAP measurements for the light (left panel), intermediate (middle panel), and heavy (right panel) DM. Dark gray regions are excluded by the WMAP (the relic abundance of DM is too big), red regions are in agreement with the WMAP results, in the remaining regions the abundance of the DM is too low. The figures come from Ref.~\cite{jhep} and~\cite{Sokolowska:2013blois}.
\label{wmap}}
\end{figure}

Since the relic abundance of the DM depends on the $\lczp$ parameter (it controls the annihilation of the DM particles through the Higgs boson to pairs of fermions or vector bosons) and the mass of the DM particle, the constraints following from its measurements can be combined with the requirement $\rg>0.7$~\cite{jhep, Sokolowska:2013blois}. We consider three cases: light, medium, and heavy DM, corresponding plots are shown in Fig.~\ref{wmap}.
For the light DM (left panel of Fig.~\ref{wmap}) the region allowed by the WMAP overlaps with the region where $\rg>0.7$ only for $M_H\gtrsim 53\g$. This means that if $H$ is supposed to account for all the observed DM, it cannot be too light. The intermediate DM can have correct relic density being in agreement with $\rg>0.7$ (middle panel of Fig.~\ref{wmap}). However, in this case $\rg$ is always suppressed with respect to the SM. The heavy DM (right panel of Fig.~\ref{wmap}) can account for both enhanced and suppressed $\rg$, being in agreement with the WMAP results, however, the allowed deviations from $\rg=1$ are very small, at the level of a few per mil.
\section{Summary}

The IDM successfully confronts present data, being in agreement with the LEP, LHC and WMAP measurements. The $h\to \gamma\gamma$ channel can provide us with important information about the model, as $\rg$ is sensitive to both $\m$ and $M_H$. Particularly interesting bounds can be obtained when also WMAP data is incorporated in the analysis.

By combining the WMAP and $\rg$ constraints we found that if substantial enhancement of $\rg$ is observed, then the charged scalar of the IDM has to be fairly light, $70\g<\m\lesssim154\g$. Moreover, the DM can only have mass in the intermediate regime and can  constitute only a fraction of the observed relic density. On the other hand, if we demand that $H$ constitutes 100\% of the DM present in the Universe, then the light DM is excluded. If $H$ is the only DM component and has intemediate mass, then $\rg$ is suppressed with respect to the SM. For the heavy DM $\rg\approx1$.


\section*{Acknowledgments}
M.K. and B.\'{S}. would like to thank the Organizers for the invitation to this nice conference. This work was supported in part by the grant NCN OPUS 2012/05/B/ST2/03306 (2012-2016).


\end{document}